\documentclass[epj,online]{webofc}
\fancypagestyle{firstpage}{%
  \fancyhf{}%
  \renewcommand{\headrulewidth}{0pt}%
  \renewcommand{\footrulewidth}{0pt}%
  \fancyhead[R]{%
    \minipage[t][\baselineskip]{\linewidth}
      \small
      XIIth Quark Confinement and the Hadron Spectrum \\
      to appear in EPJ Web of Conferences \\
    \endminipage
  }%
}
\usepackage[utf8]{inputenc}
\usepackage[varg]{txfonts}   % Web of Conferences font
\usepackage{xcolor}
\definecolor{darkred}{rgb}{0.4,0.0,0.0}
\definecolor{darkgreen}{rgb}{0.0,0.4,0.0}
\definecolor{darkblue}{rgb}{0.0,0.0,0.4}
\usepackage[bookmarks,linktocpage,colorlinks,
    linkcolor = darkred,
    urlcolor  = darkblue,
    citecolor = darkgreen]{hyperref}
\wocname{EPJ Web of Conferences}
\woctitle{CONF12}

\newcommand{\beq}{\begin{equation}}
\newcommand{\eeq}{\end{equation}}
\newcommand{\bea}{\begin{array}}
\newcommand{\eea}{\end{array}}
\newcommand{\beqa}{\begin{eqnarray}}
\newcommand{\eeqa}{\end{eqnarray}}

\newcommand{\Sec}[1]{Section~\ref{#1}}

\def\beqa{\begin{eqnarray}}
\def\eeqa{\end{eqnarray}}

\begin{document}
\selectlanguage{english}
\title{Dyons and Roberge - Weiss transition in lattice QCD}
%
% subtitle (optional, strongly discouraged)

\author{V.~G. Bornyakov\inst{1,3}\fnsep\thanks{\email{vitaly.bornyakov@ihep.ru}} \and
        D.~L. Boyda\inst{1,2,4} \and
        V.~A. Goy\inst{1,2,4} \and
        E.-M. Ilgenfritz\inst{5} \and
        B.~V. Martemyanov\inst{4} \and
        A.~V.~Molochkov\inst{1,4} \and
        Atsushi Nakamura\inst{1,6,7} \and
        A.~A. Nikolaev\inst{1,4} \and
        V.~I. Zakharov\inst{1,4}
        % etc.
}

\institute{School of Biomedicine, Far Eastern Federal University, 690950 Vladivostok, Russia
\and
           School of Natural Sciences, Far Eastern Federal University, 690950 Vladivostok, Russia
\and
           Institute for High Energy Physics NRC Kurchatov Institute, 142281 Protvino, Russia,
\and
           Institute of Theoretical and Experimental Physics NRC Kurchatov Institute, 117218 Moscow, Russia
\and
Joint Institute for Nuclear Research, BLTP, 141980 Dubna, Russia
\and
           Research Center for Nuclear Physics (RCNP), Osaka University, Ibaraki, Osaka, 567-0047, Japan,
\and
           Theoretical Research Division, Nishina Center, RIKEN, Wako 351-0198, Japan
}

\abstract{%
  We study lattice QCD with $N_f=2$ Wilson fermions
at nonzero imaginary chemical potential and nonzero temperature. We relate
the Roberge - Weiss phase transition to the properties of dyons which are
constituents of the KvBLL calorons. We present numerical evidence
that the characteristic features of the spectral gap of the overlap Dirac
operator as function of an angle modifying the boundary condition are
determined by the $Z_3$ sector of the respective imaginary chemical potential.
We then demonstrate that dyon excitations in thermal configurations
could be responsible (in line with perturbative excitations)
for these phenomena.
}
\maketitle
\section{Introduction}
\label{sec:introduction}
Confinement (of quarks and gluons) and
spontaneous breaking of chiral symmetry at low temperature and density
are two basic properties of QCD.
These properties are
connected with each other and originate from the complex structure of
the QCD vacuum state. The latter is reflected, for example, in the
condensates of gluon and quark fields.
These condensates (as vacuum expectation values) are of course space and
time independent but field fluctuations contributing to them are both
space-time and scale dependent.
Considerable activity in Lattice Gauge Theory has the aim to reveal the
corresponding structures. Semiclassical objects of QCD are since long
particular candidates to form these structures at the infrared scale.
The density and internal characteristics of semiclassical objects could
depend on external conditions (temperature, density etc.) thus providing
(or at least assisting) different phase transitions.

For long time it is known that the instanton mechanism is able to explain
chiral symmetry breaking while it was impossible to construct an instanton
mechanism for confinement in terms of an instanton gas or liquid, which are
simple realizations of a multi-instanton-antiinstanton system.
%% changed !! the simplest ?
Specific constituent dyons of Kraan-van Baal-Lee-Lu (KvBLL)
calorons~\cite{Kraan:1998pm,Kraan:1998sn,Lee:1998bb} (with twist),
however, can -- in the same way as instantons -- explain chiral symmetry
breaking. But calorons with their dyon ``substructure'' are
able also to
reproduce different features of confinement (Polyakov loop correlators,
spatial string tension, vortex and/or monopole percolation). All this has
added qualitative arguments
to the expectations already existing for decades that ``instanton quarks''
(carriers of fractional topological charge) might solve the confinement
problem.
Moreover, dyons when considered as rarefied gas (in three dimensions),
either without interaction or with some kind of Coulomb-like interaction,
give confining behavior for space-like Wilson loops and for correlators
of Polyakov loops.
This idea has been developed from the 70's to the recent
past~\cite{Polyakov:1976fu, Martemyanov:1997ks,Gerhold:2006sk,Diakonov:2007nv,Bruckmann:2011yd}.

The modelling of dyon ensembles with interaction has attracted more attention
recently~\cite{Shuryak:2011aa,Faccioli:2013ja,Larsen:2014yya,Liu:2015ufa,Liu:2015jsa,Larsen:2015vaa,Liu:2016thw}.
Therefore, it was important before as it is important now to identify dyons
in thermal lattice configurations (from quenched simulations or from full
QCD with dynamical fermions) which are thought to represent lattice gauge
fields at
different temperatures and possibly further external parameters). The aim
is to assess the relevance of these models and in particular to clarify the
importance of dyon degrees of freedom.

The caloron with nontrivial holonomy~\cite{Kraan:1998pm,Kraan:1998sn,Lee:1998bb}
has the remarkable property that the single zero mode of the Dirac operator
is located only on the ``twisted'' (Kaluza-Klein) dyon constituent when
standard antiperiodic boundary conditions are applied to the Dirac spectrum.
Depending on different temporal boundary condition (b.c.) applied to the Dirac
operator (with improved chiral properties), this zero mode may delocalize and
localize again on distinct constituent dyons~\cite{GarciaPerez:1999ux,Chernodub:1999wg}. Under certain circumstances these dyons can appear as distinct
entities, but the fractional topological charge $1/N_c$ is only realized
in the case of ``maximally nontrivial holonomy'', {\it i.e.} in the confinement
phase.

For definiteness, inspecting thermal lattice configurations of fixed total
topological charge $Q=\pm 1$ (below and above $T_c$) such a change
of the single zero mode's location with the change of b.c. was initially
observed in~\cite{Gattringer:2002wh,Gattringer:2002tg} and interpreted
according to the caloron picture taking care of the unit topological charge.

This property of mobility (and a changing degree of localization) is shared
also by a band of near-zero modes of the overlap Dirac operator as was shown
%% EMI:  Was this really clearly shown ? Where are they localized ?
%% BM: This was written in our previous work either by MMP or by you, Misha.
%% BM: We cannot ask MMP but we can ask you what have you meant by this words.
%% BM: For me it is intuitively clear that near zero modes can also jump from
%% BM: one dyon to another being more localized on heavy dyon and less on light dyon.
by some of the present authors in a series of papers
~\cite{Bornyakov:2007fm,Bornyakov:2008im,Ilgenfritz:2013oda,Bornyakov:2014esa}.

In this paper we use the above properties of calorons and of their constituent
%% EMI: Problem: Here we have no zero modes !
%% BM: In fact we have in some cases <Q^2> \neq 0. But the properties of calorons
%% BM: can be used even without their presense in confs.
dyons in order to investigate the question whether dyons can contribute to
the Roberge-Weiss (RW) phase transitions, which themselves are an outstanding
feature of simulations at imaginary chemical potential.
%% What means "contribute into" ? Do we actually mean "become manifest" ?
%% VB  I think 'contribute' is correct It is  somewhat unclear but I do
%% not know better formultion. May be 'influence'?
%% EMI: Here I agree. Let it stand !

Imaginary chemical potential is not hampered by the sign problem which, on
%% EMI: Changed
the other side, prohibts direct grand canonical simulations with real
baryonic chemical potential. Therefore standard Hybrid Monte Carlo
algorithms can be applied to simulate lattice QCD in this extension of the
case $\mu=0$. For our investigation we
employ dynamical QCD configurations primarily
generated at imaginary chemical potential as part of a project aiming to
study finite baryonic density within the canonical approach.

It is appropriate to remind the reader that there have been two reasons
to propose a detour to imaginary chemical potential, both with the final
aim to obtain (grand canonical or canonical) results for finite baryonic
density (see {\it e.g.} the review paper \cite{Muroya:2003qs}).
%%% Any newer review paper ?  G. Aarts ?

The first step in both strategies
%% frameworks
are grand canonical simulations at imaginary chemical potential.
The respective ideas are
\begin{itemize}
\item analytical continuation of observables considered as functions of
imaginary $\mu = i \mu_I$ to real $\mu$;
\item  the ``measurement'' of the grand canonical partition function as
function of imaginary chemical potential. The aim in the second case is
to make it possible to perform the Fourier transformation with respect
to $\mu_I$ in order to get the partition function for fixed baryon number
$B$ in the system (canonical approach).
\end{itemize}

These ideas have motivated the study of lattice QCD at imaginary chemical
potential in a broader sense than restricted to the two purposes.
The study of the phase transitions there (in particular the Roberge-Weiss
transition to be mentioned next) is important since their properties can
be related to properties of the phase transition in thermal QCD at zero
chemical potential (see {\it e.g.}  \cite{Bonati:2014kpa} and references
therein).

\begin{figure}[htb]
\centering
%\hspace*{-1cm}
\includegraphics[width=0.6\textwidth,angle=0]{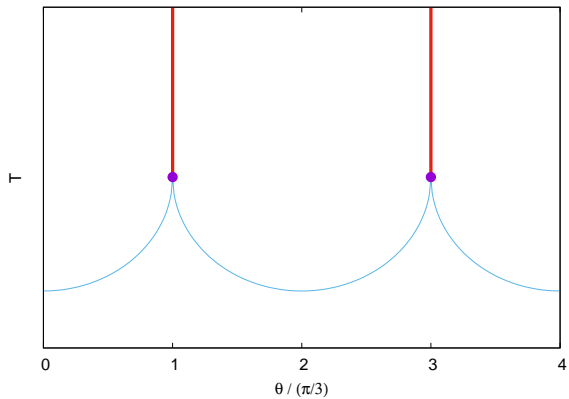}
\vspace{-0cm}
\caption{Phase diagram of QCD in the temperature--imaginary chemical potential
($\theta=\mu_I/T$) plane for intermediate quark masses. Thick vertical lines
denote first order RW transitions, thin curly lines denote crossover lines,
and thick points mark second order transition points, where a line of first
order transition ends.}
%% Does this splitting into two crossovers gives the second order points
%% a special name ?
%% VB I'm not sure
%% EMI: If there is no second order transition line to continue, I would
%% call it "second order end point", if there are three first order lines
%% meeting there, I would call it "tricritical point".
\label{fig_RW}
\end{figure}

Here we only briefly notice some basic facts of simulations with imaginary
chemical potential $\mu_q = i \mu_I$ which are important for our aim.
It is convenient to introduce an angle $\theta = \mu_I/T$ . Then the QCD
partition function at temperature $T$ is a periodic function of $\theta$:
\beq
Z(T,\theta) = Z(T,\theta+2\pi/N_c) \, .
\eeq
$N_c$ is the number of colors, in our case $N_c=3$. This periodicity property
is called Roberge-Weiss symmetry~\cite{Roberge:1986mm}.
QCD possesses a rich phase structure at nonzero $\theta$, which depends on
the number of flavors $N_f$ and the quark masses $m_q$ (or a single quark
mass in the case of $N_f = 2$).
%% VB changed:
%% This phase structure is schematically shown
The phase structure for $N_f=2$ and intermediate quark mass
is schematically shown in Fig.~\ref{fig_RW}.
%% Question: What depends here on $N_f$ ?

\begin{figure}[htb]
\centering
\hspace*{-0cm}
\includegraphics[width=0.45\textwidth,angle=270]{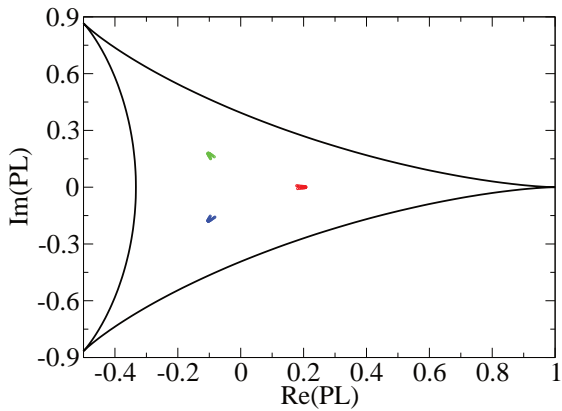}%
\vspace{-0cm}
\caption{Scatter plot for Polyakov loop $PL$ at $T = 1.35 T_c > T_{RW}$
with variation of $\theta=\mu_I/T$ inside the three basic intervals
$[-\pi/3,\pi/3]$, $[\pi/3,\pi]$, and $[\pi,5\pi/3]$
corresponding to the three $Z_3$ sectors.}
\label{pl_phase}
\end{figure}

There are first order phase transitions at $\theta = (2k+1)\pi/3$ for
temperatures $T > T_{RW}$ \cite{Roberge:1986mm} where $T_{RW}$ is somewhat
higher than $T_c$ - the temperature of the crossover to the quark-gluon
plasma  phase at zero chemical potential. These transitions are shown
as thick vertical lines in Fig.~\ref{fig_RW}.
These are transitions between $Z(3)$ sectors of the theory. A particular
sector can be identified by the phase of the average Polyakov loop.
%% VB I remove the next sentence
%% At the same time all other
%% ?? the Polyakov loop is NOT
%%averaged physical quantities are invariant under shifts of the imaginary
%chemical potential $\theta \rightarrow \theta + 2\pi/3$.
In Fig.~\ref{pl_phase} we show a scatter plot for the volume-averaged
%% EMI: changed to "volume-averaged"         over configurations
Polyakov loop $PL$ computed for different values of $\theta$.
%for an ensemble of xxx configurations. %% How many ??
%% EMI: How many in each sector ?
%% VB corrected:
The Polyakov loop has been computed for $\theta$ from 0 to  $\pi$.
For other values of $\theta$ its values were obtained
using the fact that Im$(PL)$ (Re$(PL)$) is an odd (an even) function
of $\theta$.
One can see all three patches corresponding to the three $Z_3$ sectors.

We have also checked that the values of $PL$ in one $Z_3$ sector can be
obtained by respective center transformation from its values in another
another $Z_3$ sector: $PL(\theta+2k\pi/3) = G_{k}PL(\theta)$ where
$G_k$ denotes rotation by $2k\pi/3$ in the $PL$ complex number plane.

The important role of dyons in the Roberge-Weiss transition, that we are
going to demonstrate, also underlines their importance for QCD in general.
To our best knowledge this is a first study of the role
of nonperturbative fluctuations within thermal QCD at imaginary chemical
potential. We will present evidence that the dyons could
(together with perturbative fluctuations) contribute to the transitions
and thus influence the strength of the RW transitions.
Our argument is as follows. It is evident that low modes of the quark determinant
(at least for small dynamical quark masses) play a crucial role in the resulting
phase structure depicted in Fig.~\ref{fig_RW}. It is then important to
identify the fluctuations responsible for the low modes.
We will present numerical evidence supporting idea that dyons are such
fluctuations.

In \Sec{sec:definitions2} we introduce the lattice set-up in which the
ensembles of gauge fields, that we are going to analyze have been generated
in lattice QCD with $N_f=2$ dynamical flavors and imaginary chemical potential.
We also define all the topologically relevant lattice observables employed
lateron for the analysis.

Then, in \Sec{sec:results} we present results on the spectrum.
Namely, we show the gap in the spectrum as a function of an angle in the
fermionic temporal boundary conditions for some represantative configurations
generated with different imaginary chemical potentials. Also as a counter example
we show such a gap for one configuration for $T < T_c$.

We also show gap results for artificially created configurations including
pairs of dyons and anti-dyons and argue that they have similar properties
and contribute to the partition function in such a way as to increase the
strength of the RW transition. In \Sec{sec:conclusions} we shall draw our conclusions.

%------------------------------------------------------------------------------
\section{Lattice Setting for the Thermal Ensembles}
%--------------------------------------------------
\label{sec:definitions2}

We study lattice QCD with $N_f=2$ light quarks using the clover improved
Wilson action defined by the fermion matrix
%\begin{align}
\beqa
\label{Wilson}
\Delta(n,m,\mu_q)&=&\delta_{nm} -\kappa C_{SW}\delta_{nm}\sum_{\mu\leq\nu}
\sigma_{\mu\nu}F_{\mu\nu} \notag -\kappa\sum_{i=1}^3\Big[\left(1-\gamma_i\right)U_i(n)\delta_{m,n+\hat{i}}
+\left(1+\gamma_i\right)U_i^\dagger(m)\delta_{m,n-\hat{i}}\Big] \\
&-&\kappa\Big[e^{+\mu_qa}(1-\gamma_4)U_4(n)\delta_{m,n+\hat{4}}
+e^{-\mu_qa}(1+\gamma_4)U_4^\dagger(m)\delta_{m,n-\hat{4}}\Big] \equiv1-\kappa Q(\mu_q).
\eeqa
%\end{align}
where $\kappa$ is the hopping parameter and $c_{SW}$ - the improvement
coefficient. For the gluon field we adopt the
Iwasaki improved gauge field action
\begin{equation}
S_{\mathrm{g}}=\frac{\beta}{6}\left[c_0\sum_{n,\mu<\nu}W_{\mu\nu}^{1\times1}(n)
+
c_1\sum_{n,\mu<\nu}W_{\mu\nu}^{1\times2}(n)\right]
\end{equation}
where $c_1=-0.331$ and $c_0=1-8c_1$.
We generate configurations on
$16^3 \times 4$ lattices at $\beta=2.0$, $\kappa=0.136931$ and
$c_{SW}=(1-0.8412 \beta^{-1})^{-3/4}$
%$c_{SW}=(1−0.8412\beta^{−1})^{−3/4}$.
These parameters correspond to a temperature ratio
$T/T_c=1.35$ for a quark mass determined by a target mass
ratio $m_{\pi}/m_{\rho}=0.8$.
All parameters of the action including $c_{SW}$ value were borrowed from the
WHOT-QCD collaboration paper \cite{Ejiri:2009hq}.

The overlap Dirac operator $D$ fulfills the Ginsparg-Wilson
equation~\cite{Ginsparg:1981bj}. A possible solution -- for any
{\it input Dirac operator}, in our case for the
%% added "mosstly used"
mostly used Wilson-Dirac operator
$D_W$ -- is the following zero-mass overlap Dirac
operator~\cite{Neuberger:1997fp,Neuberger:1998wv}
\beq
D(m=0)=\frac{\rho}{a}\,\left( 1 + \frac{D_W}{\sqrt{D_W^{\dagger}\,D_W}}
\right) =\frac{\rho}{a}\,\left( 1 + {\rm sgn}(D_W) \right) \,,
\label{eq:OverlapDirac}
\eeq
with $D_W = M - {\rho}/{a}$, where $M$ is the hopping term of the
Wilson-Dirac operator and ${\rho}/{a}$ is a negative mass term usually
determined by optimization.
The index of $D$, i. e. the difference of its number of right-handed and
left-handed zero modes $\psi_{0}$ with chirality $\pm 1$, can be identified
with the integer-valued topological charge
$Q_{\rm over}$ \cite{Hasenfratz:1998ri}. The non-zero modes appear in pairs, which are related to each other by $\psi_{\lambda}= \gamma_5 \psi_{-\lambda}$,
and have vanishing chirality.

The diagonalization of the overlap operator is achieved using a variant
of the Arnoldi algorithm~\cite{Neff:2001zr}. We have computed
20 lowest eigenmodes.

We consider generalized boundary conditions for any Dirac operator
underlying the diagonalization
\beq
\psi(\vec{x},x_4+\beta) = \exp(i\phi)\psi(\vec{x},x_4) \; .
\label{eq:bc1}
\eeq

While the physical fermion sea is described by the clover-improved
Wilson-Dirac operator, implemented with
antiperiodic temporal boundary conditions ($\phi=\pi$), the introduction
of the imaginary chemical potential is equivalent to
subjecting the Dirac operator for physical fermions to continuously
modified temporal boundary conditions characterized by the angle
$\theta = \mu_I/T$,
\beq
\psi(\vec{x},x_4+\beta) = \exp(i(\pi-\theta)\psi(\vec{x},x_4) \; .
\label{eq:bc1}
\eeq
%% VB phi was introduced in the Introducton
%%The following values of $\pi-\theta \equiv \phi$
The following values of $\phi$
 \beq
\phi = \left\{
\begin{array}{ll}
-\pi/3\, \\
 +\pi/3\, \\

~~~~\pi\,
\end{array}
\right\}
\label{eq:bc2}
\eeq
\noindent
correspond to elements in the boundary
condition for which on a a single caloron solution the corresponding
fermion zero modes become maximally localized at one
%% , but each time at a different one
of its three constituent dyons. Note that $\phi_3$ corresponds to the
antiperiodic boundary condition.

%----------------------------------------------------------------------
\section{Results for the spectral gap}
%% EMI: for the spectrum and the spectral gap
%----------------------------------------------------------------------
\label{sec:results}

In our study of the low modes we use the (massless) overlap lattice Dirac
operator rather than
the clover-improved Wilson-Dirac operator for which gauge field configurations
had been simulated. This is partially due to the fact that it would have exact
zero modes on topologically nontrivial configurations which helps the
safe discrimination between zero and
nonzero modes and makes possible
the unambiguous introduction of the gap into the consideration.
For brevity we will call it simply "the Dirac operator".

\begin{figure}[htb]
\centering
\includegraphics[width=0.41\textwidth,angle=270]{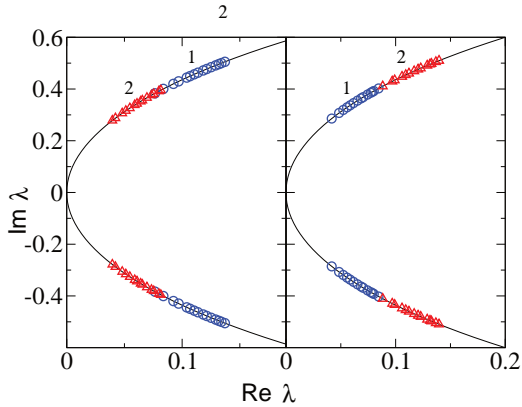}%
\vspace{0cm}
\caption{Spectra of the overlap Dirac operator for two configurations.
Left panel: two spectra
for configuration I. Spectrum 1 is for an angle $\phi$ in the overlap Dirac
operator boundary condition
$\phi_1=\phi_I $, spectrum 2 - for
$\phi_2=0.24\cdot(2\pi) < (-\theta_{II}+\pi)$.
Right panel: two spectra
for  configuration II. Spectrum 2 is for an angle $\phi$ in the overlap Dirac
operator's boundary condition
$\phi_2=\phi_{II} $, spectrum 1 - for
$\phi_1 =0.422\cdot(2\pi) > (-\theta_I+\pi)$. }
\label{fig_spectra}
\end{figure}

In Fig.~\ref{fig_spectra} we show the spectra of the Dirac operator for
two configurations.
Configuration I (left part) was generated at
$\theta_I = 1 < \pi/3$ ($\phi_I = -1 + \pi$), while
configuration II (right part) - at $\theta_{II} = 1.1 > \pi/3$
($\phi_{II}= - 1.1 +\pi$).
The configurations are taken from ensembles lying on both sides of the
first Roberge-Weiss transition value $\theta=\pi/3$.
When we change the phase $\phi$ in the overlap Dirac operator to be
diagonalized for configuration I from
%%VB changed  since 0.34 * 2pi is equal to -1+pi up to 4th digit
%% $\phi_1 = 0.34\cdot(2\pi) \approx (-\theta_I + \pi)$
$\phi_1 = \phi_I$
to $\phi_2 = 0.24\cdot(2\pi) < \phi_{II}$), we find the spectrum
changing in the left part of the figure from blue circles to red upside
triangles.
We see that even a small change of $\phi$
%% VB added:
across the phase transiton
%% VB end
%% VB  I do not understand this:
(corresponding to the displacement to the ``wrong'' phase)
%% VB end
gives rise to a drastic change of the spectrum: the
eigenvalues are shifted to smaller values thus decreasing the gap.
It is evident that this change of the spectrum implies a substantial
decreasing of the Dirac operator determinant thus indicating that the
statistical weight of configuration I displaced to the region
$\theta > \pi/3$ is small.

We make a similar observation for the configuration II after changing
the phase $\phi$ in the overlap Dirac operator from
%% VB analogous change:
%% $\phi_2=0.32\cdot(2\pi) \approx (-\theta_{II} + \pi)$
$\phi_2=\phi_{II}$
to  $\phi_1 =0.422\cdot(2\pi) > (-\theta_I + \pi)$. The respective spectra
undergo similar changes (shown in the right panel)
and thus configuration II would acquire a small statistical weight for
$\theta < \pi/3$.

We then are going to measure the spectrum for these two configurations I and
II) for a few values of $\phi$ in the Dirac operator boundary condition in
order to cover the range of $\phi$ between $0$ and $2\pi$. Additionally,
we do the same for a configuration III generated at $\theta_{III} = 3.2 > \pi$,
i.e. beyond the second Roberge-Weiss transition, in other words, in the
third $Z_3$ sector.

\begin{figure}[htb]
\centering
\includegraphics[width=0.5\textwidth,angle=270]{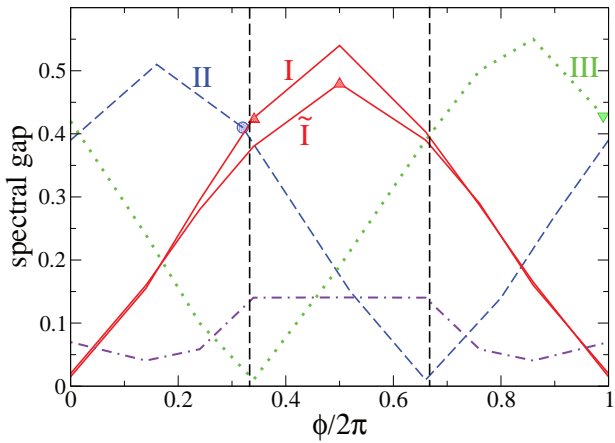}%
\vspace{0cm}
\caption{Spectral gap of the overlap Dirac operator as function of $\phi$
for four thermal
configurations (two red solid curves, one dashed blue curve and one green
dotted curve) generated at $T = 1.35 T_c$ and for one further thermal
configuration (one violet dash-dotted curve) generated at $T < T_c$.
See explanations in the text.}
\label{fig_gap1}
\end{figure}

%% EMI absatz introduced, new idea !
In Fig.~\ref{fig_gap1} we show the spectral gap as a
function of $\phi$ for configurations I (red solid lines),
II (blue dashed line) and III (green dotted line), representing all three
center sectors. The points on these curves marked with symbols correspond
to the value of $\theta$
used in the generation of the corresponding configurations. Such a dependence
of the gap on the angle in the fermionic boundary condition is already known
for different thermal configurations of SU(3) Yang-Mills theory and QCD at
zero chemical potential~\cite{Bilgici:2009tx}
but the origin of the
dependence was not discussed there.

We see from this figure that configuration I (generated at
$\phi \in [2\pi/3; 4\pi/3]$) has a maximal gap (and thus a maximal weight)
among three configurations I, II, III in the range of $\phi$ between
(approximately) $2\pi/3$ and $4\pi/3$. Configuration II (generated at
$\phi \in [0; 2\pi/3]$)  has a maximal gap in the range from 0 to $2\pi/3$,
and configuration III (generated at $\phi \in [4\pi/3; 2\pi]$) has a maximal
gap in the range from  $4\pi/3$ to $2\pi$.

Thus we observe that the Roberge-Weiss transition is accompanied by the
drastic change in the gauge field configurations: configurations generated
on one side of the transition have small statistical weight on the other
side of it due to drastic change in respective Dirac operator spectrum.

In Fig.~\ref{fig_gap1} we additionally show the spectral gap for one more
configuration $\tilde{\rm{I}}~$ from the first Roberge-Weiss sector but
generated at zero imaginary chemical potential $\theta = 0$. One can
see that the spectral gaps for two configurations generated at two different
values of $\theta$ from the same Roberge-Weiss sector are qualitatively
very similar. Thus we may extend conclusions about the spectral gaps of the
configurations I - III to all configurations of all three Roberge-Weiss
sectors.

It is important to clarify what fluctuations of the gauge field are responsible
for these
%% such
properties of the Dirac operator spectrum.

Let us now consider the Dirac operator spectrum in the background of an
artificially created configuration. Such configuration was considered
in \cite{Bornyakov:2014esa}, details of its construction can be found in
that paper. The three dyons of a given caloron have different actions
 determined by the asymptotic holonomy.
For the holonomy
%% close to
approaching one of $Z_3$ center element there are two light dyons and one heavy
dyon. We start from a caloron-anticaloron configuration and remove the heavy
dyon and the heavy antidyon. Thus, our configuration consists of two light
dyons and two light antidyons, see eq.~(36) and eq.~(37) of
Ref.~\cite{Bornyakov:2014esa}.

In \cite{Bornyakov:2014esa} we have presented numerical evidence that dyons
of the heavy type are rare in the deconfining phase of lattice QCD with zero
chemical potential.

It is known that for a caloron configuration the zero mode of the Dirac
operator is localized on one of the dyons depending on boundary conditions
(or equivalently, on the holonomy). This is also true for near zero modes.
For a heavy dyon-antydyon pair the near-zero modes appear when holonomy
is close to $Z_3=I$ and $\phi=\pi$ (antiperiodic boundary conditions),
see the right-most panel of Fig.~1 in Ref.~\cite{Bornyakov:2014esa}.
In distinction to that,
%% opposite,
when we consider a configuration without heavy (anti)dyons, e.g. a
configuration consisting of a light double dyon - antidyon pair we find
rather large gap in the spectrum, see the right-most panel of Fig.~4 in
Ref.~\cite{Bornyakov:2014esa}.
In \cite{Bornyakov:2014esa} we have presented numerical evidence that in
the deconfinement phase of lattice SU(3) gluodynamics heavy dyons
are suppressed and low modes in the spectrum of thermal configurations
are resembling the spectrum of the artificial configuration with light
double dyon - antidyon pair.

\begin{figure}[htb]
\centering
\includegraphics[width=0.41\textwidth,angle=270]{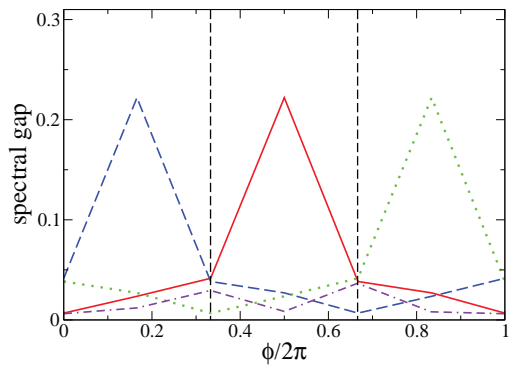}%
\vspace{0cm}
\caption{Spectral gap of the overlap Dirac operator for three artificial
configuration as function of $\phi$. See explanations in the text.}
\label{fig_double_pair}
\end{figure}

In Fig.~\ref{fig_double_pair} we show the spectral gap for the configuration
with a light double dyon - antidyon pair as function of $\phi$. We see
that the dependence of the gap on $\phi$ is qualitatively the same as that
of thermal configurations from the $Z_3=I$ Roberge-Weiss sector (red solid
curve in Fig.~\ref{fig_gap1}):
the maximum is at $\phi=\pi$ and the gap decreases to a minimal value (which
is rather close to zero) at $\phi=0$ or $\phi=2\pi$.
For artificial configurations that differ from this one by center
transformations $Z_3 = e^{i2\pi/3}$ or $Z_3 = e^{i4\pi/3}$ light and heavy
dyons simply change their enumeration
$$1\rightarrow 2, 2\rightarrow 3, 3\rightarrow 1$$
or
$$1\rightarrow 3, 3\rightarrow 2, 2\rightarrow 1$$
(see the corresponding blue dashed and green dotted curves).

Results obtained for the spectral gap for these artificial dyon configurations
demonstrate that such configurations allow to explain the main characteristic
features of the spectral gap obtained for equilibrium configurations shown in
Fig.~\ref{fig_gap1}.
It is important to notice that the results for artificial caloron-anticaloron
configuration (all three dyons and antidyons present, represented by the
violet dash-dotted curve in Fig.~\ref{fig_double_pair}) are qualitatively
different from the results for the artificial dyon configuration: the
spectral gap practically (compared to the other curves in
Fig.~\ref{fig_double_pair}) does not depend on the boundary condition.
In this respect these results are similar to the results of Fig.~\ref{fig_gap1}
where the gap for a configuration from the confining phase (violet dash-dotted
curve) has the same property.

\section{Conclusions.}
\label{sec:conclusions}

We made a study of the low modes spectrum of the overlap Dirac operator
lattice QCD with $N_f=2$ flavors of clover-improved Wilson fermions at
$T/T_c=1.35$ and $T/T_ < 1$
(with $m_\pi/m_\rho=0.8$) at nonzero imaginary chemical potential.
We have considered equilibrium configurations generated at particular
values of the imaginary chemical potential taken from all three Roberge-Weiss
(center) sectors. We have computed 20 lowest modes of the Dirac operator for
every such configuration varying the imaginary chemical potential in the
analysing Dirac operator.

We observed (see  Fig.~\ref{fig_gap1}) that the spectral gap has some
uniform characteristic features for equilibrium configuration generated
at non-vanishing imaginary chemical potential $\theta$ anywhere in the given
Roberge-Weiss (center) sector.
\begin{itemize}
\item This gap (as a function of an angle in the boundary condition for
the analysing overlap Dirac operator) has a maximum in the middle of the
Roberge-Weiss sector and decreases monotonously;
\item this behavior is independent of the particular value of $\theta$
inside the sector where the configuration has been generated.
\item The spectral gaps for two sectors intersect at the boundary between
them. This means that configurations generated at $\theta$
from one sector have low statistical weight in another sector.
\end{itemize}

We have argued that the reason for such behavior of the spectral gap could be
partially connected to nonperturbative objects present in equilibrium
configuration, constituent dyons of calorons.

{\bf Acknowledgments}\\
The authors are grateful to Ph. de Forcrand for useful remarks.
The work was completed due to support of the Russian Science Foundation
via grant number 15-12-20008. Computer simulations were performed on the FEFU GPU cluster Vostok-1 and  MSU 'Lomonosov' supercomputer.

%Journal Author, Journal \textbf{Volume}, page numbers (year)
%Book Author, \textit{Book title} (Publisher, place, year) page numbers
% etc
%\end{thebibliography}


\begin{thebibliography}{31}
\expandafter\ifx\csname natexlab\endcsname\relax\def\natexlab#1{#1}\fi
\expandafter\ifx\csname bibnamefont\endcsname\relax
  \def\bibnamefont#1{#1}\fi
\expandafter\ifx\csname bibfnamefont\endcsname\relax
  \def\bibfnamefont#1{#1}\fi
\expandafter\ifx\csname citenamefont\endcsname\relax
  \def\citenamefont#1{#1}\fi
\expandafter\ifx\csname url\endcsname\relax
  \def\url#1{\texttt{#1}}\fi
\expandafter\ifx\csname urlprefix\endcsname\relax\def\urlprefix{URL }\fi
\providecommand{\bibinfo}[2]{#2}
\providecommand{\eprint}[2][]{\url{#2}}


\bibitem[{\citenamefont{Kraan and van Baal}(1998{\natexlab{a}})}]{Kraan:1998pm}
\bibinfo{author}{\bibfnamefont{T.~C.} \bibnamefont{Kraan}} \bibnamefont{and}
  \bibinfo{author}{\bibfnamefont{P.}~\bibnamefont{van Baal}},
  \bibinfo{journal}{Nucl.Phys.} \textbf{\bibinfo{volume}{B533}},
  \bibinfo{pages}{627} (\bibinfo{year}{1998}{\natexlab{a}}),
  \eprint{hep-th/9805168}.

\bibitem[{\citenamefont{Kraan and van Baal}(1998{\natexlab{b}})}]{Kraan:1998sn}
\bibinfo{author}{\bibfnamefont{T.~C.} \bibnamefont{Kraan}} \bibnamefont{and}
  \bibinfo{author}{\bibfnamefont{P.}~\bibnamefont{van Baal}},
  \bibinfo{journal}{Phys.Lett.} \textbf{\bibinfo{volume}{B435}},
  \bibinfo{pages}{389} (\bibinfo{year}{1998}{\natexlab{b}}),
  \eprint{hep-th/9806034}.

\bibitem[{\citenamefont{Lee and Lu}(1998)}]{Lee:1998bb}
\bibinfo{author}{\bibfnamefont{K.-M.} \bibnamefont{Lee}} \bibnamefont{and}
  \bibinfo{author}{\bibfnamefont{C.-H.} \bibnamefont{Lu}},
  \bibinfo{journal}{Phys.Rev.} \textbf{\bibinfo{volume}{D58}},
  \bibinfo{pages}{025011} (\bibinfo{year}{1998}), \eprint{hep-th/9802108}.

\bibitem[{\citenamefont{Polyakov}(1977)}]{Polyakov:1976fu}
\bibinfo{author}{\bibfnamefont{A.~M.} \bibnamefont{Polyakov}},
  \bibinfo{journal}{Nucl. Phys.} \textbf{\bibinfo{volume}{B120}},
  \bibinfo{pages}{429} (\bibinfo{year}{1977}).

\bibitem[{\citenamefont{Martemyanov and Molodtsov}(1997)}]{Martemyanov:1997ks}
\bibinfo{author}{\bibfnamefont{B.}~\bibnamefont{Martemyanov}} \bibnamefont{and}
  \bibinfo{author}{\bibfnamefont{S.}~\bibnamefont{Molodtsov}},
  \bibinfo{journal}{JETP Lett.} \textbf{\bibinfo{volume}{65}},
  \bibinfo{pages}{142} (\bibinfo{year}{1997}).

\bibitem[{\citenamefont{Gerhold et~al.}(2007)\citenamefont{Gerhold, Ilgenfritz,
  and M{\"u}ller-Preussker}}]{Gerhold:2006sk}
\bibinfo{author}{\bibfnamefont{P.}~\bibnamefont{Gerhold}},
  \bibinfo{author}{\bibfnamefont{E.-M.} \bibnamefont{Ilgenfritz}},
  \bibnamefont{and}
  \bibinfo{author}{\bibfnamefont{M.}~\bibnamefont{M{\"u}ller-Preussker}},
  \bibinfo{journal}{Nucl. Phys.} \textbf{\bibinfo{volume}{B760}},
  \bibinfo{pages}{1} (\bibinfo{year}{2007}), \eprint{hep-ph/0607315}.

\bibitem[{\citenamefont{Diakonov and Petrov}(2007)}]{Diakonov:2007nv}
\bibinfo{author}{\bibfnamefont{D.}~\bibnamefont{Diakonov}} \bibnamefont{and}
  \bibinfo{author}{\bibfnamefont{V.}~\bibnamefont{Petrov}},
  \bibinfo{journal}{Phys.Rev.} \textbf{\bibinfo{volume}{D76}},
  \bibinfo{pages}{056001} (\bibinfo{year}{2007}), \eprint{0704.3181}.

\bibitem[{\citenamefont{Bruckmann et~al.}(2012)\citenamefont{Bruckmann, Dinter,
  Ilgenfritz, Maier, M{\"u}ller-Preussker, and Wagner}}]{Bruckmann:2011yd}
\bibinfo{author}{\bibfnamefont{F.}~\bibnamefont{Bruckmann}},
  \bibinfo{author}{\bibfnamefont{S.}~\bibnamefont{Dinter}},
  \bibinfo{author}{\bibfnamefont{E.-M.} \bibnamefont{Ilgenfritz}},
  \bibinfo{author}{\bibfnamefont{B.}~\bibnamefont{Maier}},
  \bibinfo{author}{\bibfnamefont{M.}~\bibnamefont{M{\"u}ller-Preussker}},
  \bibnamefont{and} \bibinfo{author}{\bibfnamefont{M.}~\bibnamefont{Wagner}},
  \bibinfo{journal}{Phys.Rev.} \textbf{\bibinfo{volume}{D85}},
  \bibinfo{pages}{034502} (\bibinfo{year}{2012}), \eprint{1111.3158}.

\bibitem[{\citenamefont{Shuryak}(2012)}]{Shuryak:2011aa}
\bibinfo{author}{\bibfnamefont{E.}~\bibnamefont{Shuryak}}, \bibinfo{journal}{J.
  Phys.} \textbf{\bibinfo{volume}{G39}}, \bibinfo{pages}{054001}
  (\bibinfo{year}{2012}), \eprint{1112.2573}.

\bibitem[{\citenamefont{Faccioli and Shuryak}(2013)}]{Faccioli:2013ja}
\bibinfo{author}{\bibfnamefont{P.}~\bibnamefont{Faccioli}} \bibnamefont{and}
  \bibinfo{author}{\bibfnamefont{E.}~\bibnamefont{Shuryak}},
  \bibinfo{journal}{Phys. Rev.} \textbf{\bibinfo{volume}{D87}},
  \bibinfo{pages}{074009} (\bibinfo{year}{2013}), \eprint{1301.2523}.

\bibitem[{\citenamefont{Larsen and Shuryak}(2014)}]{Larsen:2014yya}
\bibinfo{author}{\bibfnamefont{R.}~\bibnamefont{Larsen}} \bibnamefont{and}
  \bibinfo{author}{\bibfnamefont{E.}~\bibnamefont{Shuryak}}
  (\bibinfo{year}{2014}), \eprint{1408.6563}.

\bibitem[{\citenamefont{Liu et~al.}(2015{\natexlab{a}})\citenamefont{Liu,
  Shuryak, and Zahed}}]{Liu:2015ufa}
\bibinfo{author}{\bibfnamefont{Y.}~\bibnamefont{Liu}},
  \bibinfo{author}{\bibfnamefont{E.}~\bibnamefont{Shuryak}}, \bibnamefont{and}
  \bibinfo{author}{\bibfnamefont{I.}~\bibnamefont{Zahed}}
  (\bibinfo{year}{2015}{\natexlab{a}}), \eprint{1503.03058}.

\bibitem[{\citenamefont{Liu et~al.}(2015{\natexlab{b}})\citenamefont{Liu,
  Shuryak, and Zahed}}]{Liu:2015jsa}
\bibinfo{author}{\bibfnamefont{Y.}~\bibnamefont{Liu}},
  \bibinfo{author}{\bibfnamefont{E.}~\bibnamefont{Shuryak}}, \bibnamefont{and}
  \bibinfo{author}{\bibfnamefont{I.}~\bibnamefont{Zahed}}
  (\bibinfo{year}{2015}{\natexlab{b}}), \eprint{1503.09148}.

\bibitem[{\citenamefont{Larsen and Shuryak}(2015)}]{Larsen:2015vaa}
\bibinfo{author}{\bibfnamefont{R.}~\bibnamefont{Larsen}} \bibnamefont{and}
  \bibinfo{author}{\bibfnamefont{E.}~\bibnamefont{Shuryak}}
  (\bibinfo{year}{2015}), \eprint{1504.03341}.

\bibitem{Liu:2016thw}
  Y.~Liu, E.~Shuryak and I.~Zahed,
  %``The Instanton-Dyon Liquid Model III: Finite Chemical Potential,''
  arXiv:1606.07009 [hep-ph].

\bibitem[{\citenamefont{Garcia~Perez et~al.}(1999)\citenamefont{Garcia~Perez,
  Gonzalez-Arroyo, Pena, and van Baal}}]{GarciaPerez:1999ux}
\bibinfo{author}{\bibfnamefont{M.}~\bibnamefont{Garcia~Perez}},
  \bibinfo{author}{\bibfnamefont{A.}~\bibnamefont{Gonzalez-Arroyo}},
  \bibinfo{author}{\bibfnamefont{C.}~\bibnamefont{Pena}}, \bibnamefont{and}
  \bibinfo{author}{\bibfnamefont{P.}~\bibnamefont{van Baal}},
  \bibinfo{journal}{Phys.Rev.} \textbf{\bibinfo{volume}{D60}},
  \bibinfo{pages}{031901} (\bibinfo{year}{1999}), \eprint{hep-th/9905016}.

\bibitem[{\citenamefont{Chernodub et~al.}(2000)\citenamefont{Chernodub, Kraan,
  and van Baal}}]{Chernodub:1999wg}
\bibinfo{author}{\bibfnamefont{M.~N.} \bibnamefont{Chernodub}},
  \bibinfo{author}{\bibfnamefont{T.~C.} \bibnamefont{Kraan}}, \bibnamefont{and}
  \bibinfo{author}{\bibfnamefont{P.}~\bibnamefont{van Baal}},
  \bibinfo{journal}{Nucl. Phys. Proc. Suppl.} \textbf{\bibinfo{volume}{83}},
  \bibinfo{pages}{556} (\bibinfo{year}{2000}), \eprint{hep-lat/9907001}.

\bibitem[{\citenamefont{Gattringer}(2003)}]{Gattringer:2002wh}
\bibinfo{author}{\bibfnamefont{C.}~\bibnamefont{Gattringer}},
  \bibinfo{journal}{Phys. Rev.} \textbf{\bibinfo{volume}{D67}},
  \bibinfo{pages}{034507} (\bibinfo{year}{2003}), \eprint{hep-lat/0210001}.

\bibitem[{\citenamefont{Gattringer and Schaefer}(2003)}]{Gattringer:2002tg}
\bibinfo{author}{\bibfnamefont{C.}~\bibnamefont{Gattringer}} \bibnamefont{and}
  \bibinfo{author}{\bibfnamefont{S.}~\bibnamefont{Schaefer}},
  \bibinfo{journal}{Nucl. Phys.} \textbf{\bibinfo{volume}{B654}},
  \bibinfo{pages}{30} (\bibinfo{year}{2003}), \eprint{hep-lat/0212029}.

\bibitem[{\citenamefont{Bornyakov et~al.}(2007)\citenamefont{Bornyakov,
  Ilgenfritz, Martemyanov, Morozov, M{\"u}ller-Preussker, and
  Veselov}}]{Bornyakov:2007fm}
\bibinfo{author}{\bibfnamefont{V.}~\bibnamefont{Bornyakov}},
  \bibinfo{author}{\bibfnamefont{E.-M.} \bibnamefont{Ilgenfritz}},
  \bibinfo{author}{\bibfnamefont{B.}~\bibnamefont{Martemyanov}},
  \bibinfo{author}{\bibfnamefont{S.}~\bibnamefont{Morozov}},
  \bibinfo{author}{\bibfnamefont{M.}~\bibnamefont{M{\"u}ller-Preussker}},
  \bibnamefont{and} \bibinfo{author}{\bibfnamefont{A.}~\bibnamefont{Veselov}},
  \bibinfo{journal}{Phys.Rev.} \textbf{\bibinfo{volume}{D76}},
  \bibinfo{pages}{054505} (\bibinfo{year}{2007}), \eprint{0706.4206}.

\bibitem[{\citenamefont{Bornyakov et~al.}(2009)\citenamefont{Bornyakov,
  Ilgenfritz, Martemyanov, and M{\"u}ller-Preussker}}]{Bornyakov:2008im}
\bibinfo{author}{\bibfnamefont{V.}~\bibnamefont{Bornyakov}},
  \bibinfo{author}{\bibfnamefont{E.-M.} \bibnamefont{Ilgenfritz}},
  \bibinfo{author}{\bibfnamefont{B.}~\bibnamefont{Martemyanov}},
  \bibnamefont{and}
  \bibinfo{author}{\bibfnamefont{M.}~\bibnamefont{M{\"u}ller-Preussker}},
  \bibinfo{journal}{Phys.Rev.} \textbf{\bibinfo{volume}{D79}},
  \bibinfo{pages}{034506} (\bibinfo{year}{2009}), \eprint{0809.2142}.

\bibitem[{\citenamefont{Ilgenfritz et~al.}(2014)\citenamefont{Ilgenfritz,
  Martemyanov, and M{\"u}ller-Preussker}}]{Ilgenfritz:2013oda}
\bibinfo{author}{\bibfnamefont{E.-M.} \bibnamefont{Ilgenfritz}},
  \bibinfo{author}{\bibfnamefont{B.}~\bibnamefont{Martemyanov}},
  \bibnamefont{and}
  \bibinfo{author}{\bibfnamefont{M.}~\bibnamefont{M{\"u}ller-Preussker}},
  \bibinfo{journal}{Phys.Rev.} \textbf{\bibinfo{volume}{D89}},
  \bibinfo{pages}{054503} (\bibinfo{year}{2014}), \eprint{1309.7850}.

\bibitem[{\citenamefont{Bornyakov et~al.}(2015)\citenamefont{Bornyakov,
  Ilgenfritz, Martemyanov, and Muller-Preussker}}]{Bornyakov:2014esa}
\bibinfo{author}{\bibfnamefont{V.~G.} \bibnamefont{Bornyakov}},
  \bibinfo{author}{\bibfnamefont{E.~M.} \bibnamefont{Ilgenfritz}},
  \bibinfo{author}{\bibfnamefont{B.~V.} \bibnamefont{Martemyanov}},
  \bibnamefont{and}
  \bibinfo{author}{\bibfnamefont{M.}~\bibnamefont{Muller-Preussker}},
  \bibinfo{journal}{Phys. Rev.} \textbf{\bibinfo{volume}{D91}},
  \bibinfo{pages}{074505} (\bibinfo{year}{2015}), \eprint{1410.4632}.

\bibitem[{\citenamefont{Muroya et~al.}(2003)\citenamefont{Muroya, Nakamura,
  Nonaka, and Takaishi}}]{Muroya:2003qs}
\bibinfo{author}{\bibfnamefont{S.}~\bibnamefont{Muroya}},
  \bibinfo{author}{\bibfnamefont{A.}~\bibnamefont{Nakamura}},
  \bibinfo{author}{\bibfnamefont{C.}~\bibnamefont{Nonaka}}, \bibnamefont{and}
  \bibinfo{author}{\bibfnamefont{T.}~\bibnamefont{Takaishi}},
  \bibinfo{journal}{Prog. Theor. Phys.} \textbf{\bibinfo{volume}{110}},
  \bibinfo{pages}{615} (\bibinfo{year}{2003}), \eprint{hep-lat/0306031}.

\bibitem[{\citenamefont{Bonati et~al.}(2014)\citenamefont{Bonati, de~Forcrand,
  D'Elia, Philipsen, and Sanfilippo}}]{Bonati:2014kpa}
\bibinfo{author}{\bibfnamefont{C.}~\bibnamefont{Bonati}},
  \bibinfo{author}{\bibfnamefont{P.}~\bibnamefont{de~Forcrand}},
  \bibinfo{author}{\bibfnamefont{M.}~\bibnamefont{D'Elia}},
  \bibinfo{author}{\bibfnamefont{O.}~\bibnamefont{Philipsen}},
  \bibnamefont{and}
  \bibinfo{author}{\bibfnamefont{F.}~\bibnamefont{Sanfilippo}},
  \bibinfo{journal}{Phys. Rev.} \textbf{\bibinfo{volume}{D90}},
  \bibinfo{pages}{074030} (\bibinfo{year}{2014}), \eprint{1408.5086}.

\bibitem[{\citenamefont{Roberge and Weiss}(1986)}]{Roberge:1986mm}
\bibinfo{author}{\bibfnamefont{A.}~\bibnamefont{Roberge}} \bibnamefont{and}
  \bibinfo{author}{\bibfnamefont{N.}~\bibnamefont{Weiss}},
  \bibinfo{journal}{Nucl. Phys.} \textbf{\bibinfo{volume}{B275}},
  \bibinfo{pages}{734} (\bibinfo{year}{1986}).

\bibitem[{\citenamefont{Ejiri et~al.}(2010)\citenamefont{Ejiri, Maezawa, Ukita,
  Aoki, Hatsuda, Ishii, Kanaya, and Umeda}}]{Ejiri:2009hq}
\bibinfo{author}{\bibfnamefont{S.}~\bibnamefont{Ejiri}},
  \bibinfo{author}{\bibfnamefont{Y.}~\bibnamefont{Maezawa}},
  \bibinfo{author}{\bibfnamefont{N.}~\bibnamefont{Ukita}},
  \bibinfo{author}{\bibfnamefont{S.}~\bibnamefont{Aoki}},
  \bibinfo{author}{\bibfnamefont{T.}~\bibnamefont{Hatsuda}},
  \bibinfo{author}{\bibfnamefont{N.}~\bibnamefont{Ishii}},
  \bibinfo{author}{\bibfnamefont{K.}~\bibnamefont{Kanaya}}, \bibnamefont{and}
  \bibinfo{author}{\bibfnamefont{T.}~\bibnamefont{Umeda}}
  (\bibinfo{collaboration}{WHOT-QCD}), \bibinfo{journal}{Phys. Rev.}
  \textbf{\bibinfo{volume}{D82}}, \bibinfo{pages}{014508}
  (\bibinfo{year}{2010}), \eprint{0909.2121}.

\bibitem[{\citenamefont{Ginsparg and Wilson}(1982)}]{Ginsparg:1981bj}
\bibinfo{author}{\bibfnamefont{P.~H.} \bibnamefont{Ginsparg}} \bibnamefont{and}
  \bibinfo{author}{\bibfnamefont{K.~G.} \bibnamefont{Wilson}},
  \bibinfo{journal}{Phys. Rev.} \textbf{\bibinfo{volume}{D25}},
  \bibinfo{pages}{2649} (\bibinfo{year}{1982}).

\bibitem[{\citenamefont{Neuberger}(1998{\natexlab{a}})}]{Neuberger:1997fp}
\bibinfo{author}{\bibfnamefont{H.}~\bibnamefont{Neuberger}},
  \bibinfo{journal}{Phys. Lett.} \textbf{\bibinfo{volume}{B417}},
  \bibinfo{pages}{141} (\bibinfo{year}{1998}{\natexlab{a}}),
  \eprint{hep-lat/9707022}.

\bibitem[{\citenamefont{Neuberger}(1998{\natexlab{b}})}]{Neuberger:1998wv}
\bibinfo{author}{\bibfnamefont{H.}~\bibnamefont{Neuberger}},
  \bibinfo{journal}{Phys. Lett.} \textbf{\bibinfo{volume}{B427}},
  \bibinfo{pages}{353} (\bibinfo{year}{1998}{\natexlab{b}}),
  \eprint{hep-lat/9801031}.

\bibitem[{\citenamefont{Hasenfratz et~al.}(1998)\citenamefont{Hasenfratz,
  Laliena, and Niedermayer}}]{Hasenfratz:1998ri}
\bibinfo{author}{\bibfnamefont{P.}~\bibnamefont{Hasenfratz}},
  \bibinfo{author}{\bibfnamefont{V.}~\bibnamefont{Laliena}}, \bibnamefont{and}
  \bibinfo{author}{\bibfnamefont{F.}~\bibnamefont{Niedermayer}},
  \bibinfo{journal}{Phys. Lett.} \textbf{\bibinfo{volume}{B427}},
  \bibinfo{pages}{125} (\bibinfo{year}{1998}), \eprint{hep-lat/9801021}.

\bibitem{Neff:2001zr}
  H.~Neff, N.~Eicker, T.~Lippert, J.~W.~Negele and K.~Schilling,
  %``On the low fermionic eigenmode dominance in QCD on the lattice,''
  Phys.\ Rev.\ D {\bf 64}, 114509 (2001)
  doi:10.1103/PhysRevD.64.114509
  [hep-lat/0106016].

\bibitem[{\citenamefont{Bilgici et~al.}(2009)}]{Bilgici:2009tx}
\bibinfo{author}{\bibfnamefont{E.}~\bibnamefont{Bilgici}} \bibnamefont{et~al.}
  (\bibinfo{year}{2009}), \eprint{0906.3957}.

\end{thebibliography}
\end{document}